\documentclass[journal]{IEEEtran}

\usepackage{array}
\newcolumntype{P}[1]{>{\centering\arraybackslash}p{#1}}
\newcolumntype{M}[1]{>{\centering\arraybackslash}m{#1}}

\usepackage{url}
\usepackage{color}
\usepackage{multirow}
\usepackage{graphicx}
\usepackage{enumitem}
\usepackage{subfigure,tabularx}
\usepackage{amsfonts,amsmath,color,amssymb,amsxtra,graphicx,floatflt}
\usepackage{hhline}
\usepackage{booktabs,multirow}
\usepackage{longtable}
\usepackage{fancyhdr}

\begin{document}

\title{Smart City IoT Services Creation through Large Scale Collaboration}

%
%
%
\author{Flavio Cirillo, 
        David G\'omez,
        Luis Diez, 
        Ignacio Elicegui Maestro,
        Thomas Barrie Juel Gilbert,
        Reza Akhavan
\thanks{F. Cirillo is with: NEC Labs Europe, Heidelberg, DE; Uni. Federico II, Naples, IT. (e-mail: flavio.cirillo@\{neclab.eu,unina.it\})} 
\thanks{D. G\'omez and I. Elicegui Maestro are with ATOS Research and Innovation, Santander, ES (e-mail: \{david.gomez,ignacio.elicegui\}@atos.net)}
\thanks{L. Diez is with Uni. of Cantabria, Santander, ES, ldiez@tlmat.unican.es}
\thanks{T. Gilbert is with Alexandra Instituttet, Aarhus, DK (email: thomas.gilbert@alexandra.dk)}
\thanks{R. Akhavan is with Connected Places Catapult, London, UK, (email: reza.akhavan@cp.catapult.org.uk)}
\thanks{\copyright2020 IEEE.  Personal use of this material is permitted.  Permission from IEEE must be obtained for all other uses, in any current or future media, including reprinting/republishing this material for advertising or promotional purposes, creating new collective works, for resale or redistribution to servers or lists, or reuse of any copyrighted component of this work in other works. DOI: 10.1109/JIOT.2020.2978770}
}

\markboth{IEEE Internet of Things Journal}%
{Shell \MakeLowercase{\textit{et al.}}: Bare Demo of IEEEtran.cls for IEEE Journals}

\maketitle

\begin{abstract}
Smart cities solutions are often monolithically implemented, from sensors data handling through to the provided services. The same challenges are regularly faced by different developers, for every new solution in a new city. Expertise and know-how can be re-used and the effort shared.
In this article we present the methodologies to minimize the efforts of implementing new smart city solutions and maximizing the sharing of components. The final target is to have a live technical community of smart city application developers. The results of this activity comes from the implementation of 35 city services in 27 cities between Europe and South Korea. To share efforts, we encourage developers to devise applications using a modular approach. Single-function components that are re-usable by other city services are packaged and published as standalone components, named \textit{Atomic Services}. We identify 15 atomic services addressing smart city challenges in data analytics, data evaluation, data integration, data validation, and visualization. 38 instances of the atomic services are already operational in several smart city services.
We detail in this article, as atomic service examples, some data predictor components. Furthermore, we describe real-world atomic services usage in the scenarios of Santander and three Danish cities.
The resulting atomic services also generate a side market for smart city solutions, allowing expertise and know-how to be re-used by different stakeholders.
\end{abstract}

\begin{IEEEkeywords}
\end{IEEEkeywords}

\IEEEpeerreviewmaketitle

\section{Introduction}

Internet-of-Things (IoT) benefits in the smart city scenario is extensively acknowledged by countless real application deployments~\cite{KIM2017159}. The developments of these applications usually happen as standalone activities resulting in city-wide IoT segments fragmentation, or even domain-wide within the same city~\cite{vital2015}. 
The SynchroniCity project aims to ``synchronize''~\cite{cirillo2019} IoT infrastructures among cities in order to overcome vendor lock-in, by adopting open solutions, and city lock-in, by adopting common interfaces~\cite{oasc_mim}. The approach of SynchroniCity is to deploy an overlay on top of existing smart city infrastructures. The SynchroniCity overlay is based on FIWARE, an IoT framework of open source components implementing open API and standards.
This gives the ground to city services developers to develop applications relying on standardized interfaces and data models. The advantage is porting IoT solutions from a context to another with minimal effort, thus, enabling an IoT services market ecosystem.

The initial cities, named Reference Zones (RZs), are: Antwerp, Carouge, Eindhoven, Helsinki, Manchester, Milan, Porto, and Santander. Those cities homogenize their IoT data with the same data models~\cite{FIWARE-datamodels} and formats. This data are exposed with standard interfaces for context management (Next Generation Service Interface - NGSI~\cite{oma_ngsi}), historical time-series, and open data. 
Using these common interfaces, service developers implemented several smart city applications addressing use cases defined by local municipalities~\cite{SynCity-D3-1}. 

This article reports the collaboration activities among the applications developers (including academia, SMEs and industries). The approach is to build the applications by composing small services, namely \textit{atomic services}, each implementing a single functional block, towards Service-Oriented Architecture (SOA)~\cite{abu-matar}. The atomic services are exchanged between city application developers and published in a one-stop-shop repository\footnote{https://gitlab.com/synchronicity-iot?filter=atomic+service} following a one-page documentation scheme. The advantage of this approach is multi-fold: 1) small companies can leverage others' know-how to speed-up applications implementation, breaching the barrier of monolithic vertical developments only sustainable for big companies~\cite{AIOTI-WG02}; 2) it enables an IoT services market~\cite{biotope,westerlund2014}; 3) operational smart city applications can quickly adapt to IoT evolution by having the atomic services up to date or integrating new ones.

In this article we present the process of giving rise to 15 atomic services during 22 months (from Jan 2018 till Oct 2019) in parallel with the implementation of 35 smart city services. A bootstrap phase comprises a static analysis of use case requirements from RZs~\cite{sc_giots19} and of available off-the-shelf services. During a second phase, 12 RZs' city services architectures are analyzed to identify common challenges and building blocks. The still ongoing third phase gives the developers the possibility to spontaneously offer their IoT services as atomic services. The latter phase involves 23 city services piloted in 27 cities. In one case an application team, rather than implementing an atomic service, leveraged the know-how of the developer of situation prediction atomic services (i.e., parking and traffic flow estimator) by requesting a new one for outdoor noise estimation. This shows the marketability of developers' know-how and IoT services. In other cases, i.e. for 3 cities in Denmark (Aarhus, Vejle and Odense), the proposed atomic services aim to keep pace of the IoT evolution (e.g., the new NGSI-LD standard~\cite{ETSI-CIM-2019}) and to smooth integration for escaping city lock-in. These atomic services address data integration~\cite{Desai-SemanticGateway}, data validation, and data visualization~\cite{Khalid-IoTDataVisualization-2018}. The contribution of this article is:

\begin{itemize}[leftmargin=*]

\item \textbf{Analysis of cities commonalities}. We evaluate data input and targeting challenges of 12 services in 9 cities. The results motivate us to create a smart city technical community.

\item \textbf{Design a collaborative approach} for starting and maintaining the community alive.

\item \textbf{Atomic services examples and usage in real scenario}. We describe parking/traffic/noise estimators service as examples of atomic services. In addition, we report the experience of the real city scenarios and how they benefited by using atomic services: implement a new multi-modal transportation city service in Santander; bring legacy city services of Vejle, Aarhus and Odense to open market.

\item \textbf{Evaluate the atomic service approach} through a validation process and the assessment of the community engagement.
\end{itemize}



\section{Related Work}
\label{sec:relatedwork}

Building interoperable services on top of IoT systems is an open and timely research challenge, as demonstrated by the presence of recent works in the literature. Experimentation-as-a-Service (EaaS)~\cite{FIESTAIEEEAccess} is a paradigm to execute processing routines over a centralized platform that offers data. Re-usable and portable experiments process data regardless its origin. The architecture presented in \cite{FIESTAIEEEAccess} has the technical potential to offer experiments as re-usable services but this is not explored by the authors.
Several aspects are already tackled to enable IoT services ecosystem, such as architectures~\cite{Broering-2017,Thuluva-2017} and procedures to acquire IoT services~\cite{Thuluva-2017}, or how to generate IoT services business model~\cite{Schladofsky-2017,Ziouvelou-2020}. The authors of \cite{biotope} present a proof-of-concept of IoT services marketplace. None of the mentioned works show to be embodied in reality. The reason is that a community of such kind is not easy to self-blossom but needs to be guided.

Big cloud providers, such as Amazon Web Services~\cite{AWS-marketplace} and Microsoft Azure~\cite{Azure-marketplace}, offer service marketplaces, but mainly for industrial IoT projects. This is due to the reluctance from city governance to fall in vendor lock-in trap~\cite{Ahlgren-2016,Cirillo-FIWARE-2019}.
FIWARE, instead, is a growing open alternative to proprietary platforms~\cite{Cirillo-FIWARE-2019}. In particular, FIWARE domain-specific enablers (DSEs)~\cite{FIWARE-dse} are a collection of IoT applications and services for different domains, such as manufacturing, media, e-health, agrifood or energy. The catalogue is formed by re-usable components, similar to our atomic services, and monolithic applications. The FIWARE DSEs methodology is to simply share applications' software built upon the FIWARE framework. However, FIWARE DSEs lacks: a) a methodology to systematically identify re-usable components as services, b) a large number of IoT services and involved parties, c) the attempt to keep the community alive. Hence, our work is complementary to the FIWARE DSEs.

\section{Smart city services}
\label{sec:atomicservices}

\begin{table*}[t!]
\caption{SynchroniCity smart city services}
\label{tab:smartcityservices}
\centering
\resizebox{\textwidth}{!}{%
\begin{tabular}{|P|P|P|P|P|P|}

\hline

\multicolumn{2}{|c|}{\begin{tabular}{@{}c@{}}Smart City Theme\end{tabular}} & 
\multicolumn{1}{c|}{\begin{tabular}{@{}c@{}}City Service Topic\end{tabular}} &
\multicolumn{1}{c|}{\begin{tabular}{@{}c@{}}N. of\\services\end{tabular}} &
\multicolumn{1}{c|}{\begin{tabular}{@{}c@{}}Involved Cities\end{tabular}} &
\multicolumn{1}{c|}{\begin{tabular}{@{}c@{}}N. of\\pilots\end{tabular}} \\

\hline

\multicolumn{1}{|l|}{\multirow{3}{*}{\rotatebox{90}{\hspace{-3.5em}Mobility}}} & 
\multicolumn{1}{l|}{\begin{tabular}{@{}l@{}}Encouraging non-moto-\\rized transport\end{tabular}} &
\multicolumn{1}{l|}{\begin{tabular}{@{}l@{}}insightful (clean air, crowdsourced, safe) bicycle path\\recommendation; secure bike parking; stolen bike recovery;\\electric bike usage monitoring\end{tabular}} &
\multicolumn{1}{c|}{8} &
\multicolumn{1}{l|}{\begin{tabular}{@{}l@{}}Eindhoven (NL), Milan (IT), Antwerp (BE), Santander,\\La Nuc\'{i}a(ES), Manchester (UK), Dublin, Donegal (IE),\\Faro (PT), Helsinki (FI), Aarhus (DK)\end{tabular}} &
\multicolumn{1}{c|}{18} \\

\cline{2-6}

\multicolumn{1}{|l|}{}& 
\multicolumn{1}{l|}{\begin{tabular}{@{}l@{}}Multi-modal\\transportation\end{tabular}} &
\multicolumn{1}{l|}{\begin{tabular}{@{}l@{}}commuter assistant; park \& ride; public transportation\\usage maximization; zero emission journey planner;\\barrier-free planner for disabled people;
\end{tabular}} &
\multicolumn{1}{c|}{6} &
\multicolumn{1}{l|}{\begin{tabular}{@{}l@{}}Porto (PT), Santander (ES), Helsinki (FI), Milan (IT),\\Carouge (CH), Seongnam (KR)\end{tabular}} &
\multicolumn{1}{c|}{6} \\
\cline{2-6}

\multicolumn{1}{|l|}{}& 
\multicolumn{1}{l|}{\begin{tabular}{@{}l@{}}Enabling Mobility\\as a Service (MaaS)\end{tabular}} &
\multicolumn{1}{l|}{\begin{tabular}{@{}l@{}}adaptive lighting; traffic optimization; bus stops crowd and\\air monitoring; smart parking\end{tabular}} &
\multicolumn{1}{c|}{5} &
\multicolumn{1}{l|}{\begin{tabular}{@{}l@{}}Porto (PT), Santander, Torrelavega (ES), Milan (IT),\\Antwerp (BE), Seongnam (KR)\end{tabular}} &
\multicolumn{1}{c|}{10} \\

\hline

\multicolumn{1}{|c|}{\multirow{3}{*}{\rotatebox{90}{\hspace{-3em}Sustainability}}} & 
\multicolumn{1}{l|}{\begin{tabular}{@{}l@{}}Climate Change\\Adaptation\end{tabular}} &
\multicolumn{1}{l|}{\begin{tabular}{@{}l@{}}green roof management; building energy management
\end{tabular}} &
\multicolumn{1}{c|}{3} &
\multicolumn{1}{l|}{\begin{tabular}{@{}l@{}}Carouge (CH), Milan (IT), Eindhoven (NL), Porto (PT),\\Antwerp (BE), Vejle, Odense (DK)\end{tabular}} &
\multicolumn{1}{c|}{8} \\

\cline{2-6}

\multicolumn{1}{|l|}{}& 
\multicolumn{1}{l|}{\begin{tabular}{@{}l@{}}Reducing Air and\\Noise Pollution\end{tabular}} &
\multicolumn{1}{l|}{\begin{tabular}{@{}l@{}}indoor/outdoor air quality management; clean air around\\schools; noise pollution planning; urbanization impact\\monitoring
\end{tabular}} &
\multicolumn{1}{c|}{5} &
\multicolumn{1}{l|}{\begin{tabular}{@{}l@{}}Helsinki, Tampere (FI), Santander, Bilbao, Onda (ES),\\Antwerp (BE), Carouge (CH), Edinburgh (UK),\\Novi Sad (RS), Eindhoven (NL)\end{tabular}} &
\multicolumn{1}{c|}{13} \\

\cline{2-6}

\multicolumn{1}{|l|}{}& 
\multicolumn{1}{l|}{Waste Management}& 
\multicolumn{1}{l|}{\begin{tabular}{@{}l@{}}waste collection optimization; waste collection monitoring\end{tabular}} &
\multicolumn{1}{c|}{2} &
\multicolumn{1}{l|}{\begin{tabular}{@{}l@{}}Porto (PT), Carouge (CH), Catalayud (ES), Aarhus (DK)\end{tabular}} &
\multicolumn{1}{c|}{4} \\

\hline

\multicolumn{1}{|l|}{\multirow{2}{*}{\rotatebox{90}{Governance}}}& 
\multicolumn{1}{l|}{\begin{tabular}{@{}l@{}}Community Policy\\Suite\end{tabular}} &
\multicolumn{1}{l|}{\begin{tabular}{@{}l@{}}agile governance; data visualization; public spaces air and\\noise monitoring; insights for cycling infrastructure;\\smart city business intelligence; disable people accessibility\\monitoring; traffic insights; environment monitoring; bus\\stops crowd and air monitoring; elderly care service monitoring\end{tabular}} &
\multicolumn{1}{c|}{13} &
\multicolumn{1}{l|}{\begin{tabular}{@{}l@{}}Manchester, Edinburgh(UK), Porto (PT), Carouge (CH),\\Cabildo de la Palma, Cartagena, Santander, Bilbao,\\Torrelaveda, Onda (ES), Eindhoven (NL), Antwerp (BE),\\Milan (IT), Helsinki(FI), Aarhus, Vejle, Odense (DK) \end{tabular}} &
\multicolumn{1}{c|}{23} \\

\cline{2-6}

\multicolumn{1}{|l|}{}& 
\multicolumn{1}{l|}{\begin{tabular}{@{}l@{}}Increasing citizen\\engagement in\\decision making\end{tabular}} &
\multicolumn{1}{l|}{\begin{tabular}{@{}l@{}}ease open data accessibility; open data visualization;\\citizens engagements on urbanization\end{tabular}} &
\multicolumn{1}{c|}{3} &
\multicolumn{1}{l|}{\begin{tabular}{@{}l@{}}Porto (PT), Manchester (UK), Santander (ES), Milan (IT),\\Herning (DK), Antwerp (BE), Carouge (CH),\\Novi Sad (RS), Helsinki (FI)\end{tabular}} &
\multicolumn{1}{c|}{11} \\

\hline

\multicolumn{2}{|l|}{Data Mining}& 
\multicolumn{1}{l|}{\begin{tabular}{@{}l@{}}data lake value extraction\end{tabular}} &
\multicolumn{1}{c|}{1} &
\multicolumn{1}{l|}{\begin{tabular}{@{}l@{}}Carouge (CH), Bordeaux (FR), Seongnam (KR)\end{tabular}} &
\multicolumn{1}{c|}{3} \\

\hline

\multicolumn{2}{|l|}{Privacy}& 
\multicolumn{1}{l|}{\begin{tabular}{@{}l@{}}citizens awareness of IoT\end{tabular}} &
\multicolumn{1}{c|}{2} &
\multicolumn{1}{l|}{\begin{tabular}{@{}l@{}}Antwerp (BE), Manchester (UK), Dublin (IE), Carouge (CH)\end{tabular}} &
\multicolumn{1}{c|}{4} \\

\hline

\end{tabular}%
}
\vspace{-1.5em}
\end{table*}

\begin{figure}[h]
\centering
\includegraphics[width=\linewidth,trim={0cm 0cm 0cm 0cm},clip]{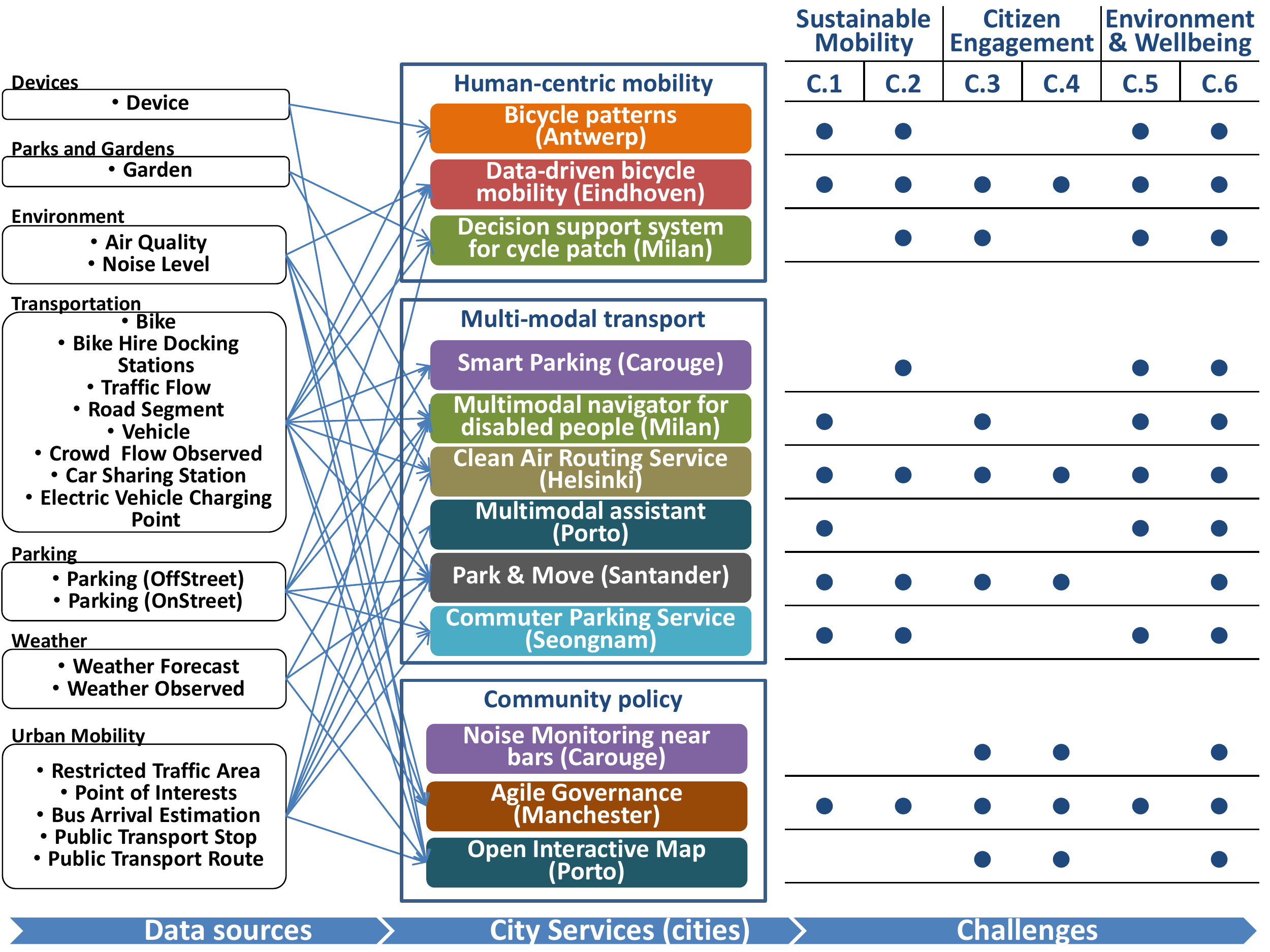}
\caption{Associations between data sources to city services, and, then, to applications challenges. Services for the same city are in boxes of same color.}
\label{fig:commonalities}
\vspace{-1em}
\end{figure}

The SynchroniCity project brings together several stakeholders with the aim of building a technical smart city community. This paper is presenting the work done towards establishing a community for smart city services. We encompass a total of 35 smart city services\footnote{https://synchronicity-iot.eu/cities-pilots/} that involve 27 different cities distributed in Europe and South Korea, reaching 72 running pilots. The themes includes (see Table~\ref{tab:smartcityservices}): 1) mobility; 2) sustainability; 3) governance; 4) data mining; 5) privacy. 

\subsection{Cities and Pilots commonalities}

Cities have latent commonalities regarding city challenges to target and available datasets to handle and process. In this section we analyse the initial 12 smart city services planned by the 8 RZs plus Seongnam (Korea). The analysis outcomes demonstrates that cities face similar challenges when developing smart services even if of different themes.
The city services are grouped by three application themes~\cite{SynCity-D3-1}: 1) human-centric mobility, 2) multi-modal transportation; 3) community policy support. Fig.~\ref{fig:commonalities} shows the associations between data sources to city services, and, then, to applications challenges. The data source types are grouped following the FIWARE data models~\cite{FIWARE-datamodels}. The challenges are:

\begin{itemize}[leftmargin=*]
 \item \textit{C.1 - Enabling Mobility-as-a-Service (MaaS)}. 
 Cities seek to shift from transportation ownership models to a service model. Targets are to enable multimodal transportations, to open real time data on transport modes, to improve efficiency of existing infrastructure, and to redistribute stakeholder roles in the ecosystem.
 
 \item \textit{C.2 - Encouraging non-motorised (active) transport}.
 Air pollution in urban areas is a known problem and cities are implementing different solutions to address it. An approach is to encourage citizens to use zero emission alternatives for short distance urban trips.
 
 \item \textit{C.3 - Increasing citizen engagement in decision making}.
 Often the decision on urban policies are left to governance without great involvement of the real beneficiaries: the citizens. Thus, even if cities are spending effort to become smart, the citizens do not perceive the benefits. Citizens involvement in the process is a new form of democracy~\cite{Mellouli-2014}.
 
 \item \textit{C.4 - Increasing diversity in political engagement}.
 Engagement to the decision making process is often viable only to whom can physically attend. This is not possible to citizens lacking time (e.g., those with family responsibilities), or who cannot easily move (e.g., elderly persons). Digitalizing the process encourages participation from whom is often silent.
 
 \item \textit{C.5 - Climate Change Adaptation}.
 Extreme urban climate shifts are recurrent all around the globe. Cities are studying different solutions to mitigate dangerous situations, such as flash flooding or extreme urban heat.

 \item \textit{C.6 - Reducing Air and Noise Pollution}
 Urban environment pollution puts citizens at different risks, such as respiratory issues (due to air pollution) or stress (due to noise pollution~\cite{Goines2007NoisePA}). Reducing or managing in an optimal way these two factors can increase life quality.

\end{itemize}

\noindent Fig.\ref{fig:commonalities} demonstrates that cities are not alone in their problems, even for different application themes. Thus, a technical community is desirable.


\section{Collaborative approach}
\label{sec:collaboration} 

To cope with a big number of data producers, service providers and cities, SynchroniCity designed a reference architecture~\cite{Cirillo-FIWARE-2019}. It has as founding principles the avoidance of \textit{city lock-in} and \textit{vendor lock-in}: the first is to ease city service replication among cities; the second is to keep the market open. The strategy is to follow the Open and Agile Smart Cities (OASC) Minimum Interoperability Mechanisms (MIMs)~\cite{oasc_mim}, to aim at a shared ecosystem of data and services. 
Operational smart city deployments are homogenized with the overlay SynchroniCity middleware, based on the open source FIWARE framework~\cite{FIWARE}, and data are exposed with common interfaces and data models~\cite{FIWARE-datamodels}. The data is then consumed by city services. 
To boost service providers collaboration and ignite the community, we steer the city services development around the concept of \textit{atomic service}. Developer teams are encouraged to build the city services following a modular paradigm, embracing the concepts of Service Oriented Architecture (SOA). If a service sub-component is generic enough to be re-usable by another city service, then it is proposed as atomic service: a single functional block consuming data and implementing any kind of feature, such as managing, enriching, joining or filtering the input. 
Atomic service has similarity with the concept of microservice in the fact of being a self-contained piece of software targeting a specific task. Nevertheless, an important characteristic is that atomic services must be re-usable. Therefore, while an atomic service instance can be a microservice, vice versa is not always true.

\begin{figure}[h]
\centering
\vspace{-1em}
\includegraphics[width=1\linewidth,trim={4cm 9.85cm 1cm 0cm},clip]{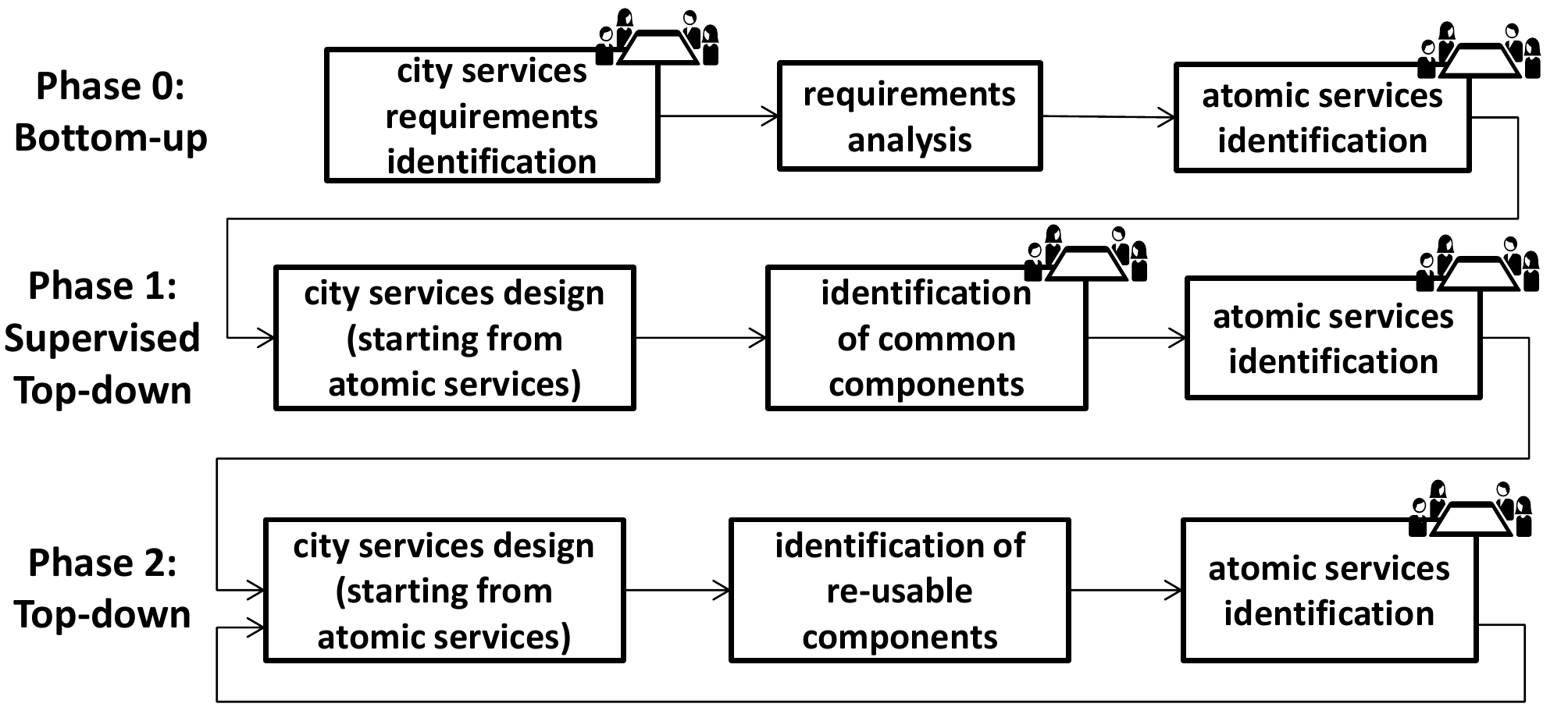}
\caption{Collaborative approach towards the implementation of 35 city services. Some of the steps (with meeting icon) involve interaction with the community.}
\label{fig:process}
\vspace{-1em}
\end{figure}

Atomic services are identified during three phases (see Fig.~\ref{fig:process}): bottom-up, a supervised top-down, and a top-down phase. The \textit{bottom-up phase} is a preamble phase to identify atomic services before city services are designed. During this period, initial city service themes (i.e., multi-modal transportation, human-centric mobility, and community policy suite) are described with use-cases and requirements with a shared effort among multiple stakeholders (i.e., municipalities, data providers, service providers, technology providers)~\cite{SynCity-D3-1}. The requirements are, then, defined by means of a questionnaire. The synthesis from all the answered questionnaires~\cite{sc_giots19} brings the identification of an initial set of atomic services. The questionnaire aims to: 1) prioritize application requirements for each city; 2) identify available re-usable software components; 3) identify available know-how by developer teams. The prioritization of requirements gave us the target to be addressed collaboratively and, thus, by an atomic service. Whereas, the listing of available software components and know-how identified re-usable software and developer teams. During the following phase, \textit{supervised top-down phase}, developer teams are left to design their city service taking into account the identified atomic services. The resulting city services' architectures are, then, jointly analyzed and compared. Common modules are selected as atomic services. 
The \textit{top-down phase} gives the field to the city services owner to freely design their own service and identify possible components that might be good candidates to become atomic services. The community, then, decides whether accepting those by checking if compliant with at least one of the following requirements: 1) to be part of two or more city services; 2) to be generic enough to have the potential to be used by multiple parties; 3) to have the potential to become part of a greater ecosystem (e.g., a Grafana plugin).
The last phase is continuous and followed by any new city service project. The first two phases lasted 13 months (from January 2018 to January 2019) and involved 12 city services in 9 cities. The last phase is still ongoing and covers other 23 city services piloted in all the 27 cities.


Once an atomic services is acknowledged, it shall comply to technical principles to ensure quality: open source code access (e.g., Gitlab), complete (and ``templated'') documentation (e.g., Apiary and ReadTheDocs), and easy to deploy (i.e., Docker).

\section{Atomic Services}
\label{sec:roster}

At the moment of writing this article (October 2019) the roster accounts 15 atomic services. Table~\ref{tab:roster} summarizes the services by category and by the phase of selection. The first three services entering the roster are the results of the bottom-up phase. During this phase, partners with previous experience in smart city projects bring their expertise and previous results. The \textit{Smart Cities Dashboard} and the \textit{Grafana Dashboard} are visualization tools used in past pilots. The \textit{Routing Service (Open Trip Planner - OTP)} is a data evaluation service, which is adopted by many current city services concerning a journey. Even if Grafana and OTP are third party software, they are still accounted as atomic services. Our approach is not to forcefully create new components, but to share best practices and expertise to city services developers.
Indeed the final goal is to create a self-sustained community. In the case of those two services, what is provided is the support, tutorial, and ready-to-use packaging for smart city context.

During the supervised top-down phase, different smart city services' architectures are collaboratively analyzed and compared. This resulted in five atomic services: \textit{Parking} and \textit{Traffic Estimator} are data analytics services that use artificial intelligence (AI) to predict, respectively, parking and traffic situation; \textit{GTFS-RT Loader}, \textit{NGSI Urban Mobility to GTFS Adapter} and \textit{GTFS Fetcher} are data integration services necessary to integrate the routing service, that digests General Transit Feed Specification (GTFS) files, with the underlying Next Generation Service Interface (NGSI) protocol adopted by the FIWARE-based framework.

During the top-down approach, other seven atomic services entered the community. The \textit{(Outdoor) Noise Estimator} is a data analytics service based on the same system of the other two estimators. This predictor resulted important for the Carouge's community policy suite service, and we have been asked to tailor a new atomic service for their needs. The \textit{Bike Data Visualiser} and \textit{Grafana NGSI Map plugin} are two visualization atomic services. The latter is a plugin for Grafana (also published in the Grafana community) to readily use NGSI data. The \textit{Transformer GPS to NGSI Traffic Flow Observed (TFO)}, and \textit{Legacy to NGSI Transformer} adapt simple JSON data to the FIWARE data models, enabling the usage of NGSI-based atomic services. The \textit{NGSI to NGSI-LD} is meant to keep the pace of the NGSI evolution towards the linked data~\cite{ETSI-CIM-2019}. Finally the \textit{NGSI Validation} checks the correctness of NGSI message against the FIWARE data models JSON schemas, increasing city services reliability.

\begin{table}[htbp]
\caption{Atomic Services selected with different approaches: phase 0) bottom-up; phase 1) supervised top-down; phase 2) top-down.}
\label{tab:roster}
\vspace{-1em}
\begin{center}
\setlength\tabcolsep{1pt}
\begin{tabular}{|c|M{12pt}|M{12pt}|M{12pt}|M{12pt}|M{12pt}|M{12pt}|M{12pt}|M{12pt}|M{12pt}|M{12pt}|M{12pt}|M{12pt}|M{12pt}|M{12pt}|M{12pt}|}

\hline
\textbf{Category} &  
\multicolumn{3}{c|}{\begin{tabular}{@{}c@{}}data\\analytics\end{tabular}} & 
\multicolumn{1}{c|}{\begin{tabular}{@{}c@{}}data\\eval.\end{tabular}} & 
\multicolumn{6}{c|}{\begin{tabular}{@{}c@{}}data\\integration\end{tabular}} & 
\multicolumn{1}{c|}{\begin{tabular}{@{}c@{}}data\\val.\end{tabular}} & 
\multicolumn{4}{c|}{\begin{tabular}{@{}c@{}}visualization\end{tabular}} \\

\hline

\begin{tabular}{@{}c@{}}\textbf{Atomic}\\\textbf{Service}\end{tabular} & 
\rotatebox{90}{parking estimator} & 
\rotatebox{90}{traffic estimator} & 
\rotatebox{90}{noise estimator} & 
\rotatebox{90}{routing} & 
\rotatebox{90}{gtfs-rt loader} & 
\rotatebox{90}{gtfs fetcher} & 
\rotatebox{90}{ngsi to gtfs} & 
\rotatebox{90}{gps to gtfs} & 
\rotatebox{90}{ngsi to ngsi-ld} & 
\rotatebox{90}{json to ngsi} & 
\rotatebox{90}{ngsi validation} & 
\rotatebox{90}{dashboard} & 
\rotatebox{90}{grafana} & 
\rotatebox{90}{bike data visualizer} & 
\rotatebox{90}{ngsi grafana plugin} \\

\hline

\textbf{Phase 0} & 
 & 
 &
 & 
\checkmark & 
 & 
 & 
 & 
 &
 & 
 & 
 & 
\checkmark & 
\checkmark & 
 & \\

\hline

\textbf{Phase 1} & 
\checkmark & 
\checkmark &
 & 
 & 
\checkmark & 
\checkmark & 
\checkmark & 
 &
 & 
 & 
 & 
 & 
 & 
 & \\

\hline

\textbf{Phase 2} & 
 & 
 &
\checkmark & 
 & 
 & 
 & 
 & 
\checkmark &
\checkmark & 
\checkmark & 
\checkmark & 
 & 
 & 
\checkmark & 
\checkmark\\

\hline

\end{tabular}
\end{center}
\end{table}

\subsection{Data analytics atomic services}

In~\cite{sc_giots19}, we introduced the legacy solution for predicting parking area availability and traffic flow based on time-series data. These services expose HTTP interfaces for providing data prediction based on machine learning (ML). 
From these two atomic services we assemble a third estimator for outdoor noise level. Namely, we leverage data from outdoor deployed noise sensors (e.g. Carouge, Santander) to predict the noise level in the next time window (e.g. 1 hour).

\begin{figure}[tbp]
    \centering
    \includegraphics[width=\columnwidth]{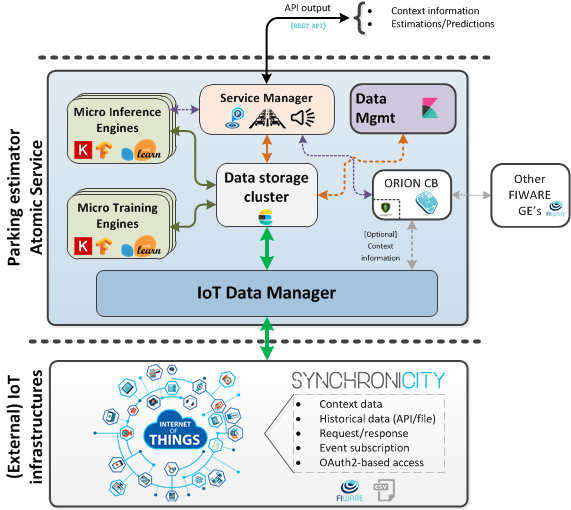}
    \caption{Estimator atomic service(s) architecture}
    \label{fig:estimator}
\end{figure}

Fig.~\ref{fig:estimator} shows the current architecture that is easier to configure/deploy/use from the previous version~\cite{sc_giots19} and, at the same time, more scalable, flexible and stable.
Following the data flow, underlying IoT infrastructures (i.e., FIWARE-based Synchronicity Framework) feed the entry point of the atomic service, so called \textit{IoT Data Manager}. This component gathers data from the various nodes/IoT devices in three ways: last values-context information, historical data, and event-based subscription. Past, present and future data is forwarded to the \textit{Data Storage Cluster}, a decentralized datastore based on Elasticsearch. 
Optionally, a FIWARE Context Broker\footnote{https://fiware-orion.readthedocs.io/} is used to store the output of the estimators as new attributes for each sensor, enabling these services to be used as standalone solutions when combined with other components (e.g., Grafana for visualization, Quantum Leap\footnote{\url{https://smartsdk.github.io/ngsi-timeseries-api/}} for timeseries storage).

The AI operations are performed by two components based on Keras, TensorFlow and Scikit-Learn as ML engine:

\begin{itemize}[leftmargin=*]
    \item \textbf{Micro Training Engine(s)} generate the models upon the data (i.e. historical time series) collected at the Data Storage Cluster. A model is generated for every context entity (i.e., a thing) having a minimum number of samples (configurable; by default, 1000). Since new data is continuously streamed onto the system, the trainer is scheduled to periodically re-train the models (by default, in a daily basis). The trainers can be configured with different algorithms, training window size, and train/test ratio.
    \item \textbf{Micro Inference Engine(s)}. For each sensor, given the model generated by a training engine and a sample chunk containing the latest observations, the inference engine calculates the predictions/estimations for the next time-window. There is one micro inference engine per micro training engine. The inference is performed, by default, every 15 minutes. The prediction values are saved into the Data Storage Cluster and, optionally, in the context broker.
\end{itemize}

Original sensor observations and estimations are exposed by two interfaces. \textit{Data Management} is a Kibana-based dashboard that visualizes data available in the Data Cluster Storage. The \textit{Service Manager} provides an HTTP API where users can: 1) query all the information stored in the Data Cluster Storage; 2) call a prediction from a Micro Inference Engine and retrieve the output. The service manager is the sole component actually different between estimators (i.e. parking, traffic and outdoor noise), since each of the Service Managers is tailored according to its respective data API and output format.

\subsection{Atomic services as building blocks for city services: multi-modal transportation city service in Santander}
\label{sec:parkandgo}

In this section we describe the combination of atomic services to provide a routing service in a multi-modal transportation scenario based on Open Trip Planner (OTP). We focus on the routing service deployed in the city of Santander that aims to provide routes within the city and from the city to nearby villages. The transportation and mobility alternatives include public bus service managed by a private company, mobility facilities within the city, which include mechanical stairs and funicular, and a private ferry service that connects the city with two villages through the bay. 

\begin{figure}
	\includegraphics[width=\columnwidth]{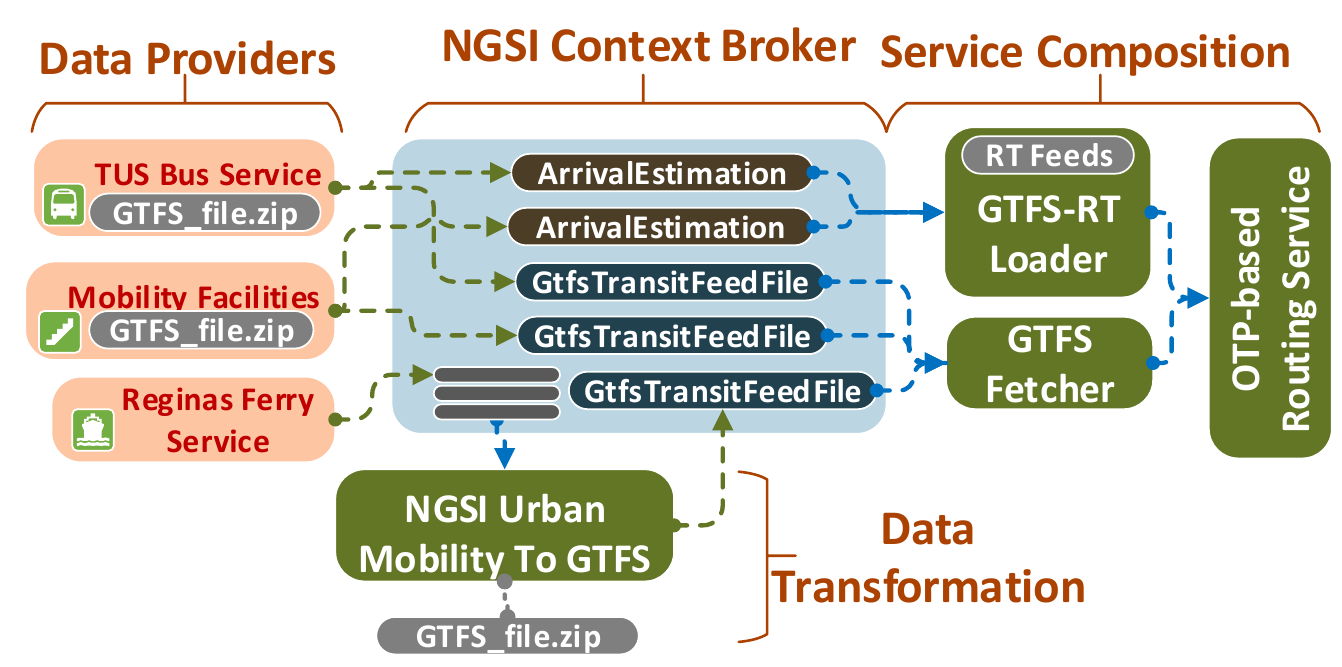}
	\vspace{-2em}
	\caption{Deployment and composition of atomic service to provide multi-modal transportation routing service in the city of Santander\label{fig:multimodal}}
	\vspace{-1em}
\end{figure}

The resulting service uses the FIWARE Context Broker as the central piece, which stores heterogeneous urban data according to the defined data models (see Fig.~\ref{fig:multimodal}). In addition, the routing engine relies on the GTFS and GTFS Real Time (GTFS-RT) data models to generate routes. 
Different data providers (city services or utilities) generate NGSI entities, that are consumed and transformed by the atomic services, so that it can be consumed by the routing engine.
It is worth noting that the public services (i.e., buses, escalators, funiculars) create entities linking to regular GTFS files using the \emph{GtfsTransitFeedFile} data model and periodically provide arrival estimation information by updating \emph{ArrivalEstimation} entities (e.g., bus arrival at a stop). On the other hand, the private service (i.e., ferries) lacks standard data models, thus, it uses a set of NGSI entities to store scheduling information.

The atomic services deployed in this scenario are:

\begin{itemize}[leftmargin=*]
	\item \textbf{NGSI Urban Mobility to GTFS} consumes NGSI urban mobility entities and generates GTFS feeds (i.e., \textit{.zip} files) storing them locally. Then, it creates GtfsTransitFeedFile entities in the context broker pointing at GTFS files. 
	
	\item \textbf{GTFS Fetcher} feeds OTP with GTFS files. This atomic service tracks the modifications in the GtfsTransitFeedFile entities to update and reload the OTP databases and maps.
	
	\item \textbf{GTFS-RT Loader} consumes ArrivalEstimation entities from the NGSI context broker, to generate GTFS-RT file. The service subscribes to ArrivalEstimation updates in the context broker, to propagate them to GTFS-RT feeds. The service also exposes this real time information through a REST interface.

	\item \textbf{Routing Service}, based on Open Trip Planner, consumes GTFS and GTFS-RT data and provides multi-modal routes, supporting customization options. The GTFS Fetcher pushes GTFS feeds to the routing service. The routing service can also pro-actively consume GTFS-RT from an endpoint. 
\end{itemize}

\subsection{Danish smart cities experiences}
Atomic services come also from the adaptation of old solution to new technologies and standards. Danish cities have been working to establish successful smart city solutions for at least a decade. Early attempts by larger municipalities are made by launching open data platforms and working with the concept of digital marketplaces where public data would be made available to foster innovation with local businesses. Later attempts include working on living labs and establishing a technical foundation for deploying IoT solutions. However, the truth is that none of this has yet made real impact, and it has not been able to scale above limited trials.

Danish cities are increasingly joining forces through networks (e.g., OS2, Gate21), to share knowledge and ensure that future smart city solutions are interoperable. Many cities have experienced pilots that have not delivered what was promised, and so their business cases have not been realized. With the market moving ever faster and global players like Microsoft and Amazon also pushing their solutions to cities, some cities feel there are too many standards, not enough national guidance and some are close to giving up entirely or capitulating to global vendors with full stack solutions.

Some danish cities like Vejle, Aarhus and Odense are working hard to break this deadlock, and these cities are using an agile approach that allows them move forward and innovate at a higher speed, and with a lower cost. Both Vejle and Odense have started to build their own simple IoT solutions from the ground up, based on their own concrete needs. Without over engineering their solutions, they have been able to build very simple ``platforms'' that can ingest data from multiple heterogeneous legacy systems into a modern ICT infrastructure. The problem is that these solutions are not based on standards, they cannot scale to meet new demands and each city is working independently of each other. Fortunately we could integrate their architecture with SynchroniCity, and all the effort they have spent understanding and parsing raw source data is not wasted. SynchroniCity project foresees that providing an excellent technical user experience for developers is key to adoption, especially in smaller cities. Individual software engineers often choose the component that enables them to deliver functionality with least effort, hence why Vejle choose to use Node-RED, influxdb and SQLserver. 

We shift Aarhus custom platform to a standard compliant (i.e., NGSI) platform using the \textit{Legacy to NGSI Transformer} atomic service that translates generic JSON format to NGSI. On the city service we replace: custom visualizations and webportals with the \textit{NGSI Grafana plugin}, and custom Redis data store with Quantum Leap. We also enhance reliability by the adopting the \textit{NGSI Validation} atomic service. In addition, we already implemented the \textit{NGSI to NGSI-LD} atomic service as preparation for the NGSI transition to linked data~\cite{ETSI-CIM-2019}. 


\section{Evaluation}
\label{sec:evaluation}

While the process of identification and publication of new atomic services is still ongoing, we validated the quality of the eight bottom-up and supervised top-down services. 
In the process, we also analyze the community engagement.

\subsection{Validation}

Synchronicity atomic services are distributed with documentation regarding the deployment process, provided functionalities (e.g. the consumed/produced datasets) and their APIs. In this sense, the validation focus is set on three main aspects: 1) documentation regarding requirements (e.g., software and hardware), installation process and usability; 2) deployment and integration with the SynchroniCity framework; 3) the features covering city service themes functional requirements. The latter point makes sense only for atomic services chosen under supervised phases, while designing city services per theme. For atomic services chosen during the top-down phase, this validation step cannot be verified since the features relate to single city services and not to city service themes.

We define a set of compositions, called scenarios, that relate atomic services based on consumed data sources and provided outcomes. These scenarios require an integration with a SynchroniCity core infrastructure to collect data. Each of these compositions evaluate the provided documentation that describes the involved atomic services, the integration with SynchroniCity framework, and their interoperability capabilities. We set up three scenarios:

\begin{itemize}[leftmargin=*]
    \item The \textbf{Routing} scenario combines GTFS information feeds with the SynchroniCity OTP-based routing service to provide multi-modal routes. It validates the capabilities of the atomic services framework to support GTFS files management and GTFS-RT feeds generation to build multi-modal routes within the city. The layout of this scenario corresponds to the Santander application (see \S\ref{sec:roster}) and puts together NGSI, GTFS and OTP services on top of the Santander SynchroniCity framework.
    
    \item The \textbf{Estimation} scenario exploits data from ParkingSpot, OnStreetParking and TrafficFlowObserved entities provided by the Santander deployment, both last value and historical data, to feed the Parking and Traffic estimators services.  
    Both services combined offer an overview of the current and incoming traffic situation on a city, providing also valuable information to upper mobility services or applications.
    
    \item  The third scenario, \textbf{City Data Visualization}, is composed by the Smart City Dashboard and Grafana Dashboard atomic services that provide views of the available IoT data and their impact on different city indicators. These two services consume data from the NGSI context broker.
\end{itemize}

\begin{figure}[ht]
\centering
\includegraphics[width=\columnwidth]{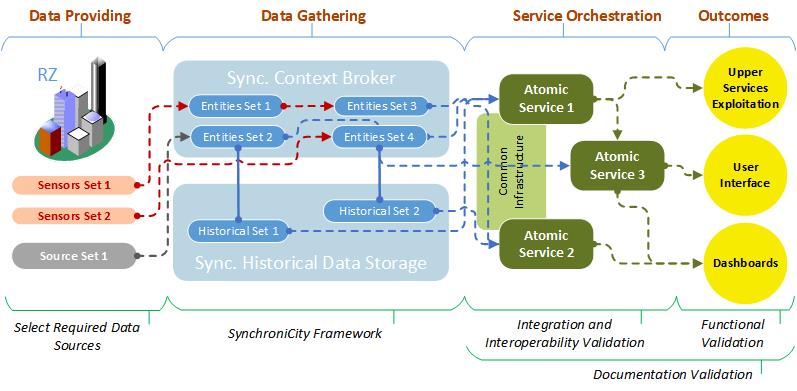}
\vspace{-2em}
\caption{Validation layout}
\label{fig:layout}
\end{figure}

We emulate a SynchroniCity core framework with a testing environment (see Figure~\ref{fig:layout}) that includes components and data sources. Following their documentation, we deploy the atomic services that directly consume SynchroniCity resources. Then, we deploy and check the rest of atomic services that consume output from the first ones (and if applicable, also from the SynchroniCity components). Finally, we validate the documented functionalities of each atomic service. This is an iterative process, thus, each detected issue in any of the validation steps, including provided documentation, is reported to the service developer to allow improvements in the next iteration.

\begin{figure}[ht]
\centering
\includegraphics[width=\columnwidth]{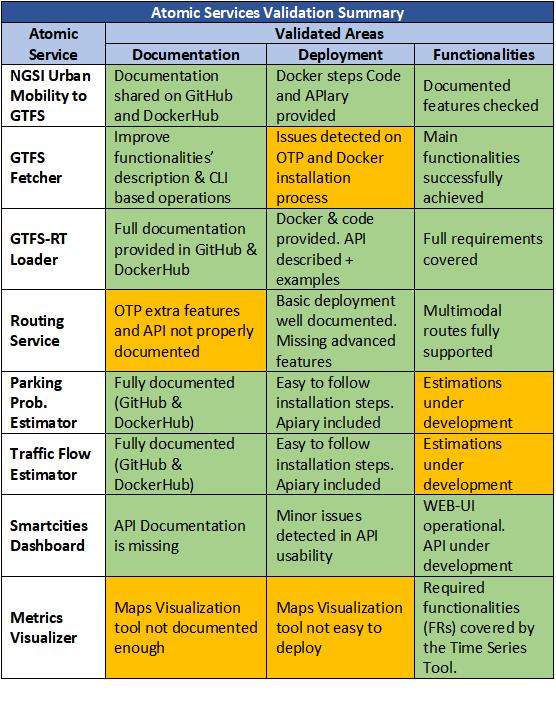}
\vspace{-3em}
\caption{Validation summary of SynchroniCity atomic services (May 2019)}
\vspace{-1em}
\label{fig:validation}
\end{figure}

Figure~\ref{fig:validation} shows a status summary shared with atomic services developers or, in case of external third party components (e.g., Grafana and OTP), the designed responsible. 

\subsection{Community Engagement}
\label{sec:results}

\begin{figure}[ht]
\centering
\includegraphics[width=\linewidth,trim={0cm 9cm 2.5cm 1cm},clip]{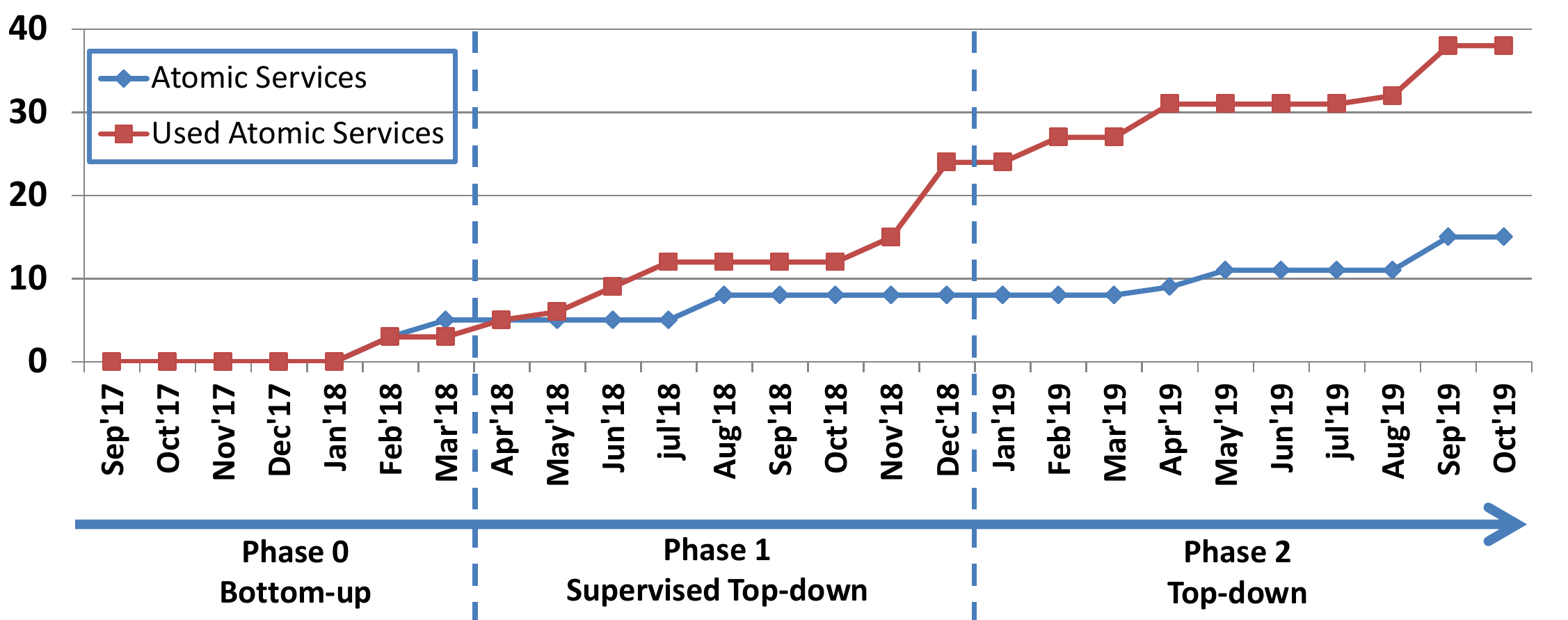}
\caption{Trend of atomic services and their adoption by city services}
\label{fig:trend}
\end{figure}

The number of atomic services increased consistently throughout time, reaching a total of 15 atomic services. Fig.~\ref{fig:trend} shows also the trend of their adoption by city services, reaching a total of 38 ``success stories''. Each integration of an atomic service in a city service is counted (e.g., the multi-modal transportation city service of Santander described in \S\ref{sec:parkandgo} counts 4 adopted atomic services).

\begin{figure}[ht]
\centering
\vspace{-1em}
\includegraphics[width=\linewidth,trim={0cm 10.5cm 2cm 0cm},clip]{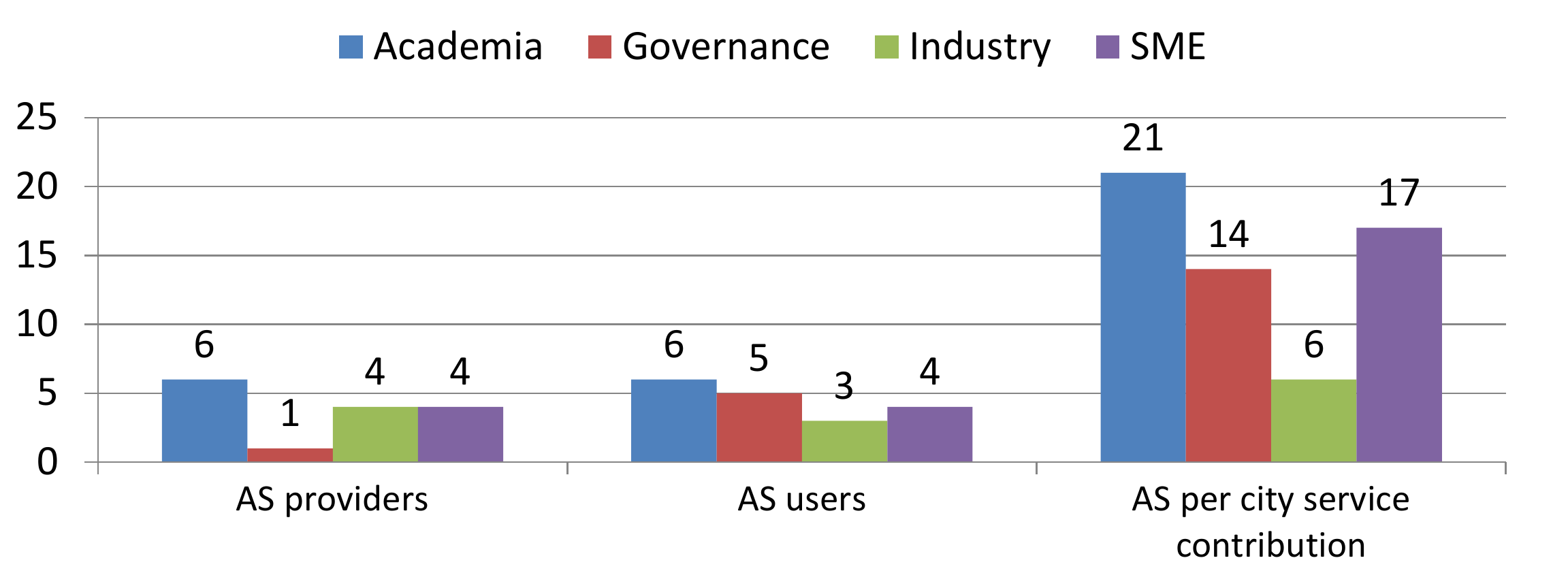}
\vspace{-2em}
\caption{Distribution of atomic service contribution and usage per stakeholder}
\label{fig:contribution}
\end{figure}

Atomic service providers are fairly distributed among the three categories of academia, industry, and SMEs (see Fig.~\ref{fig:contribution}). Among the atomic services users (i.e., parties involved in the development of city services that integrate an atomic service), also the governance category have similar share with the others. Typically, governance parties have not much resources for development. This explains why their contribution to atomic services is limited while the adoption is fostered.
If we count the parties as many times as they are involved in a city service (third set of bars in Fig.~\ref{fig:contribution}), we can observe that SMEs and governance take big advantage to this shared effort. Being academia usually the most willing to share and to experiment, this category of parties result to be the biggest in terms of providers and users. Industry also helped with 4 atomic services, but the low number of adoption is due to their involvement into a  limited number of city services.

\begin{table}[htbp]
\caption{Third party services adoption}
\vspace{-1em}
\label{tab:ownandexternal}
\begin{center}
\setlength\tabcolsep{4pt}
\begin{tabular}{|c|c|c|c|c||c|}
\hline
 & \textbf{Academia} & \textbf{Governance} & \textbf{Industry} & \textbf{SME} & \textbf{tot} \textbf{} \\ 
 \hline
 \hline

Own Adoption & 6 & 1 & 1 & 10 & 18 \\ \hline
External Adoption & 7 & 6 & 3 & 4 & 20 \\ \hline
\end{tabular}
\end{center}
\vspace{-2em}
\end{table}

\begin{table}[htbp]
\caption{City services categorizations}
\vspace{-1em}
\label{tab:servicestypology}
\begin{center}
\setlength\tabcolsep{4pt}
\begin{tabular}{|l|c|c|}
\hline
 & \textbf{City Services} & \textbf{Adopted AS} \\ 
 \hline
\hline
Public governance projects & 12 & 26 \\ \hline
6-months pilots & 16 & 0 \\ \hline
Individual exploitations & 8 & 12 \\ \hline
\end{tabular}
\end{center}
\vspace{-1em}
\end{table}

Atomic services are adopted almost evenly by the same developer party and by third parties. Table~\ref{tab:ownandexternal} shows these numbers together with the detailed by stakeholder category. It is interesting to note that while academia group is open to share and adopt ready solutions, governance and SMEs have opposite attitudes. In particular governance re-uses third parties software whereas SMEs are keen to implement their own service and offer it in the market. This is explained by the fact that startups develop their core technology to build product differentiation. This is demonstrated also in Table~\ref{tab:servicestypology} where the city services of Table~\ref{tab:smartcityservices} are categorized in public governance projects, 6-months pilots driven by startups, and individual exploitation of a single stakeholder. We note that startups project do not adopt any atomic service, mostly due to the short time project spent to develop their core technology. Cities projects, instead, are confirmed to will to re-use technologies. These two attitudes are  complementary and the atomic services community helps providers to meet customers.

If we analyze the distribution of atomic services categories by city services categories we can notice that the multi-modal transportation is the category with the biggest number of atomic services adopted. That is because the city services in this category were the most deeply designed with common effort~\cite{sc_giots19} across the bottom-up and the supervised top-down phases. Furthermore this kind of services involved advanced data manipulation (i.e., parking and traffic prediction analytics) and evaluation (i.e., routing service), with necessary data integration (i.e., to and from GTFS).
Community policy suite services take advantages from atomic services, mainly from the visualization typology (especially Grafana) being this kind of applications oriented to aid human decisions.

\begin{figure}[ht]
\centering
\vspace{-1em}
\includegraphics[width=\linewidth,trim={0cm 9.5cm 2cm 0cm},clip]{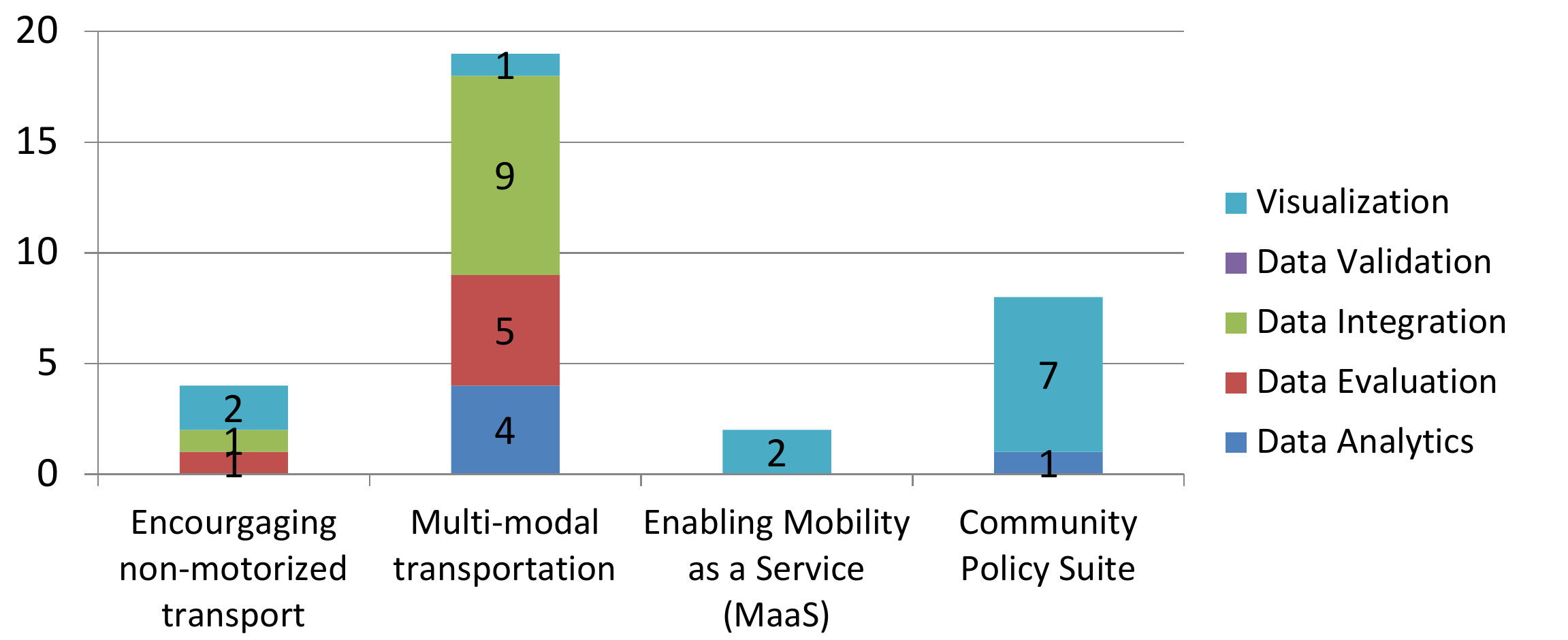}
\vspace{-2em}
\caption{Distribution of atomic service categories among city service categories}
\label{fig:servicecategories}
\end{figure} 



\section{Conclusions}
\label{sec:conclusions}

In this article we presented the collaboration among many stakeholders for the implementation of 35 city services piloted in 27 cities. The approach is to encourage developers to have modular design of their applications in order to identify re-usable IoT services, named atomic services. The identification of atomic services went through three phases, with one still open since it envisions contribution from the open community of smart city developers. We identified 15 atomic services addressing IoT challenges in data analytics, data evaluation, data integration, data validation, and visualization.
We present examples of atomic services (data estimators) and their usage in real application in Santander and Denmark. We performed validation on the quality of the atomic services and an assessment of the adoption by smart city application. It resulted that a total of 38 instances of different atomic services are operating in several city services.

Despite the igniting project (i.e., SynchroniCity) turning to an end (Dec. 2019) the atomic service community is planned to stay alive in the coming years. Therefore the atomic services community will most likely grow according to the planned exploitation by the academia, industries and SMEs.

\section*{Acknowledgment}

This work has been partially funded by the European Union's Horizon 2020 Programme under Grant Agreement No. 732240 SynchroniCity (Delivering an IoT enabled Digital Single Market for Europe and Beyond). The content of this paper does not reflect the official opinion of the European Union. Responsibility for the information and views expressed therein lies entirely with the authors.

\ifCLASSOPTIONcaptionsoff
  \newpage
\fi

\bibliographystyle{IEEEtran}
{\footnotesize
\bibliography{Bib}
}

\end{document}